\documentclass[twocolumn,showpacs,preprintnumbers,amsmath,amssymb]{revtex4}

\usepackage[dvips]{graphics}
\usepackage{amsmath, amsthm, amssymb} 
\usepackage{graphicx}
\usepackage{dcolumn}
\usepackage{bm}

\newcommand{\bra}[1]{\langle #1|}
\newcommand{\ket}[1]{|#1\rangle}

\newtheorem{proposition}{Proposition}

\begin{document}


\title{Quantifying Entanglement with Witness Operators}

\author{Fernando G. S. L. Brand\~ao}
\email{fgslb@ufmg.br}
\affiliation{Grupo de Informac\~ao Qu\^antica - Universidade Federal de Minas Gerais - Departamento  de F\'{\i}sica\\
Caixa Postal 702 - Belo Horizonte - MG -  Brasil - 30.123-970}
\date{\today}


\begin{abstract}
We present an unifying approach to the quantification of entanglement based on entanglement witnesses, which includes several already established entanglement measures such as the negativity, the concurrence and the robustness of entanglement. We then introduce an  infinite family of new entanglement quantifiers, having as its limits the best separable approximation measure and the generalized robustness. Gaussian states, states with symmetry, states constrained to super-selection rules and states composed of indistinguishable particles are studied under the view of the {\it witnessed entanglement}. We derive new bounds to the fidelity of teleportation $d_{min}$, for the distillable entanglement $E_{D}$ and for the entanglement of formation. A particular measure, the PPT-generalized robustness, stands out due to its easy calculability and provides sharper bounds to $d_{min}$ and $E_{D}$ than the negativity in most of states. We illustrate our approach studying thermodynamical properties of entanglement in the Heisenberg XXX model.
\end{abstract}

\maketitle

\section{INTRODUCTION}

In recent years entanglement has been recognized as a physical resource central to quantum information processing. As a result, a remarkable research effort has been devoted to classifying and quantifying it. The first achievement in this direction was the identification of the entropy of entanglement \cite{bennett1}, $E_{E}$, as the unique measure of entanglement for pure bipartite states in the asymptotic limit. It was shown that that $m$ copies of a pure state $\ket{\psi}$ can be reversibly converted into $n$ copies of $\ket{\phi}$ by local operations and classical communication (LOCC) if, and only if, $mE_{E}(\ket{\psi}) = nE_{E}(\ket{\phi})$. This reversibility is lost however when one considers the more general picture of mixed states. In this case two different entanglement measures, associated with the formation and distillation processes, respectively, have to be taken into account. On one hand the entanglement cost, $E_{C}(\rho)$ \cite{bennett1}, is the minimal number of singlets necessary to create the state $\rho$ by LOCC in the asymptotic regime. On the other, the distillable entanglement, $E_{D}(\rho)$ \cite{bennett1}, is the maximum number of singlets that can be extracted by LOCC from $\rho$. Another important measure connected to asymptotic properties is the relative entropy of entanglement, $E_{R}$ \cite{vedral1}. It is related to how distinguishable an entangled state is from a separable one and gives bounds to $E_{C}$ and $E_{F}$. 
 
The finite copy case is more complex and the entropic quantities considered above are not applicable anymore. For bipartite pure states, where the reversibility is already lost, the minimum set of entanglement measures characterizing deterministic and probabilistic transformations were derived \cite{nielsen1,vidal1,jonathan1}. The mixed case, however, is known only for very restricted situations and remains mainly unsolved. 

Another approach for the quantification of entanglement is to measure the usefulness of a  state to perform a given quantum information task. For example, the maximal fidelity of teleportation achieved by single copy LOCC \cite{horodecki1}, the maximal secret-key rate attainable by local measurements in a cryptographic protocol \cite{curty1} and the capacity of dense coding \cite{bruss1}, despite not being equal to any of the measures discussed so far, are clearly the best quantifiers when one of these protocols is analyzed.

Entanglement in multi-partite systems exhibits a much richer structure than the bipartite case and its study is even more challenging. Already in the pure three qubits case, there are two different manners for a state to be entangled, in the sense that there are states that cannot be converted, even with a certain probability, in each other \cite{dur1}. From the measures considered above only $E_{R}$ is unambiguously defined to multi-partite systems, although it is not the only one.

It is thus clear that entanglement is a highly complex phenomenon, which cannot be quantified by only one measure. Then, a natural way to measure it is to use any quantie which satisfies some particular properties, being the monotonicity under LOCC the most important \cite{vidal2,vedral1}. In this axiomatic approach any measure which does not increase, on average, under LOCC, called an entanglement monotone \cite{vidal2}, is a good measure of entanglement and, conversely, any meaningful quantifier has to be an entanglement monotone, or at least has some sort of weaker monotonicity under LOCC.    

A closely related problem to the quantification problem is the characterization of entanglement. The very fundamental question whether a given mixed state is entangled or not is extremely difficult, being actually NP-HARD \cite{gurvits}. A possible approach is then to consider sufficient criteria for entanglement, such as the Peres-Horodecki \cite{peres} and the alligment \cite{chen} tests. Nonetheless, the strongest manner to characterize entanglement is using entanglement witnesses (EW) \cite{horodecki2,terhal1}. They are Hermitian operators whose expectation value is positive in every separable state. Therefore, a negative expectation value in a measurement of a witness operator in an arbitrary state is a direct indication of entanglement in this state. Furthermore, it was shown by the Horodeckis that a state is entangled if, and only if, it is detected by an EW \cite{horodecki2}. A great deal of research has been devoted to the study of EWs, varying from the their classification and optimization \cite{lewenstein1,terhal2,terhal1} to their use in the characterization of entanglement in important, even macroscopic, physical systems \cite{toth1,brukner1,wu1}. Also optimal set-ups for local measurements of witnesses \cite{toth2,guhne1} and experimental realizations of witnessing entanglement were realized \cite{bou}. In spite of the determination of EWs for all states being also computationally intractable \cite{brandao1}, different methods from convex optimization theory can be applied to the problem, leading to efficient approximative procedures to determine and even optimize EWs for arbitrary states \cite{brandao1,doherty1,eisert1}. 
 
The main objective of this paper is to show that EWs can be very helpful also in the quantification of entanglement. The first measure related to EWs was due to Bertlmann {\it et al} \cite{bert} and was shown to be equal to the Hilbert-Schmidt distance from the set of separable states. Brandao and Vianna \cite{brandao2} took another significant step in this direction, showing that a measure derived from optimal EWs of the most studied group of witnesses so far, the group of EWs with unit trace, was in fact equal to the random robustness, which led to the establishment of properties still unknown for the later, such as its monotonicity under separable trace-preserving superoperators. 

Besides the obvious benefit of increasing the number of entanglement measures known, EWs based quantifiers are particular interesting due to the possibility of performing experimental measurements of them, which could be important to the extension of entanglement to other areas of physics, such as thermodynamics and statistical mechanics. Moreover, despite being necessary in general a complete tomography of a state to the determination of its degree of entanglement based on an EW measure, any EW provides a lower bound to it, even when no information at all about the state is available.               

The paper is structured as follows. In Sec. II we briefly review the basic properties of multi-partite optimal entanglement witnesses. In Sec. III we define a class of entanglement measures based on EWs, which includes several important already known quantities such as the negativity and the concurrence, and introduce a new infinite family of entanglement monotones having the generalized robustness and the best separable approximation measure as its limits. In Sec IV we present further properties of the considered measures and relate them to the localizable entanglement. In Sec. V it is shown how the methods developed in the last years to the characterization of entanglement based on convex optimization can be used to calculate approximately a large number of measures based on EWs. In Sec. VI possible extensions of our approach to Gaussian states are discussed. In Sec. VII we consider how the measures and their calculation are modified in states with symmetries. In Sec. VIII the questions of the amount of entanglement and of nonlocality in the presence of a super-selection rule are answered from the perspective of the studied measures. In Sec. IX it is shown that the three most successful approaches to the quantification of entanglement in systems of indistinguishable particles can be easily accessed from the EWs based quanties. In Sec. X and XI the questions of bounds on the teleportation distance and on the distillable entanglement of a given quantum state is review using our measures. It is shown that they provide sharper bounds than the negativity for the majority of states. In Sec. XII we derive lower bounds to the entanglement of formation with any EW. Possible applications of the measures are exemplified in Sec. XIII, where the derivation of two thermodynamic \textit{equations of state} which take into account entanglement is presented. Finally, in Sec. XIV we summarize our results and discuss future perspectives.   

\section{MULTIPARTITE SYSTEMS AND OPTIMAL ENTANGLEMENT WITNESSES}

We consider a system shared by $N$ parties ${\cal f} A_{i} {\cal g}_{i=1}^{N}$. Following \cite{dur2}, we call a $k-$partite split a partition of the system into $k \leq N$ sets ${\cal f} S_{i} {\cal g}_{i=1}^{k}$, where each may be composed of several original parties. Given a density operator $\rho_{1...k} \in {\cal B}(H_{1} \otimes ... \otimes H_{k})$ (the Hilbert space of bounded operators acting on $H_{1}\otimes ... \otimes H_{k}$) associated with some $k-$partite split, we say that $\rho_{1...k}$ is a $m$-separable state if it is possible to find a convex decomposition for it such that in each pure state term at most $m$ parties are entangled among each other, but not with any member of the other group of $n - m$ parties. For example, every 1-separable state, also called fully-separable, can be written as:
\begin{equation}
\rho_{1...k} = \sum_{i}p_{i} \ket{\psi_{i}}_{1}\bra{\psi_{i}}\otimes ... \otimes \ket{\psi_{i}}_{k}\bra{\psi_{i}}.
\end{equation}
Another example is the 2-separable states of a 3-partite split given by:
\begin{equation}
\rho_{1:2:3} = \sum_{i}p_{i}\rho_{i},
\end{equation} 
where each $\rho_{i}$ is separable with respect to at least one of the three possible partitions (A:BC, AB:C and AC:B). For each kind of separable state there is a different kind of entanglement associated to it. We will say that a state is $(m+1)$-partite entangled if it is not $m$-separable. It is clear that if a state is $m$-separable it cannot be $n$-entangled for all $n > m$. 

It is possible to detect ($m$+1)-partite entanglement using entanglement witnesses. In order to do that, consider the index set $P = {\cal f}1, 2, ..., k{\cal g}$. Let $P^{m}$ 
be a subset of $P$ which has at most $m$ elements. Then $W$ is a $(m+1)$-partite entanglement witness if
\begin{center}
\begin{equation}
\begin{array}{c}
 _{P^{m}_{v}}\bra{\psi}\otimes ... \otimes \hspace{0.07 
cm}_{P^{m}_{1}}\bra{\psi}W\ket{\psi}_{P^{m}_{1}} \otimes ... \otimes 
\ket{\psi}_{P^{m}_{v}} \geq 0 \\ \\
\forall \hspace{0.2 cm} P^{m}_{1}, ..., P^{m}_{v} \hspace{0.2 cm} $such 
that$ \\ \\
\bigcup_{k=1}^{v}P^{m}_{k} = P \hspace{0.2 cm} $and$ \hspace{0.2 cm} 
P^{m}_{k} \bigcap P^{m}_{l} = {\cal f}{\cal g}. 
\end{array}
\end{equation}
\end{center}
Equation $(3)$ assures that the operator $W$ is positive for all 
$m$-separable states. Thus, as the subspace of $m$-separable density operators is convex and closed, a state $\rho$ is $(m+1)$-entangled if and only if there is a Hermitian operator satisfying equation (3) such that $Tr(W\rho) < 0$ \cite{horodecki3}. 
        
Usually one is interest in a selected group of witnesses operators called {\it optimal}. Two different definitions of optimal entanglement witness (OEW) exist. The first, introduced by Lewenstein {\it et al} \cite{lewenstein1}, is based on how much entangled states a given entanglement witness (EW) $W$ is able to detect: $W$ is optimal iff there is no other EW which detects all the states detected by $W$ and some other states not detected by $W$. The second definition, due to Terhal \cite{terhal2}, establish the concept of OEW relative to a chosen entangled state $\rho$. The $\rho-$optimal entanglement witness $W_{\rho}$ is given by
\begin{equation}
Tr(W_{\rho}\rho) = \min_{W \in {\cal M}} Tr(W\rho),
\end{equation}         
where ${\cal M}$ is the intersection of the set of entanglement witnesses, denoted by ${\cal W}$, with some other set ${\cal C}$ such that ${\cal M}$ is compact \cite{foot1}. Note that every $\rho$-OEW is also an OEW accordingly to the first definition, whereas the converse may not be true. 

A general expression for entanglement witnesses was presented in \cite{lewenstein2}. Every EW acting on $k$-partite Hilbert space can be written as:
\begin{equation}
W = P + \sum_{i=1}^{k}Q_{i}^{T_{i}} - \epsilon I,
\end{equation}   
where $P$ and the $Q_{i}$'s are positive semi-definite, $\epsilon \geq 0$, $I$ is the identity operator and $T_{i}$ is the partial transposition with respect to partie $i$. Note that even ($m$+1)-partite EWs can be written in the form of equation (5) \cite{foot2}. An important class of EW is the decomposable entanglement witnesses (d-EW), which can always be written as: 
\begin{equation}
W = P + \sum_{i=1}^{k}Q_{i}^{T_{i}}. 
\end{equation} 
This class will be particularly important in our discussion, since the set of entangled states detected by d-EW is invariant under LOCC \cite{doherty2}.
 
\section{Definitions and basic properties}

In this section we show how $\rho$-optimal EWs can be used to quantify all the different kinds of multipartite entanglement. First, an unifying approach, which includes several important entanglement measures (EM), will be presented. Then we will consider a new infinite family of entanglement monotones \cite{vidal2}. 

A general expression for the quantification of entanglement via EWs is defined as:
\begin{equation}
E(\rho) = \max {\cal f}0, -\min_{W \in {\cal M}} Tr(W\rho) {\cal g},
\end{equation} 
where ${\cal M} = {\cal W} \cap {\cal C}$, and the set ${\cal C}$ is what distinguish the quantities. We call {\it witnessed entanglement} any measure expressed by equation (7). 

Some well known EM can be expressed as (7). The first, introduced by Bertlmann {\it et al} \cite{bert}, is:
\begin{equation}
B(\rho) = \max_{||W - I||_{2} \leq 1}[\min_{\sigma \in {\cal S}}Tr(W\sigma) - Tr(W\rho)],
\end{equation}
where $W \in {\cal W}$. $B(\rho)$ was shown to be monotonic decreasing under mixing enhancing maps \cite{hayden1} and to be equal to the ${\cal H}_{s}$-distance of $\rho$ to the set ${\cal S}$ of fully-separable states:
\begin{equation}
B(\rho) = D(\rho) = \min_{\sigma \in {\cal S}}||\rho - \sigma||_{2}.
\end{equation}

The second is the negativity, i.e., the sum of the negative eigenvalues of $\rho^{T_{A}}$ (the partial transpose of $\rho$ with respect to  subsystem A) \cite{vidal3,eisert2,zyc}. It is easily seen that ${\cal N}$ can be written as:

\begin{equation}
{\cal N}(\rho) = \max {\cal f} 0, - \min_{0 \leq W \leq I}Tr(W^{T_{A}}\rho) {\cal g}.
\end{equation}

Another quantie is the maximal fidelity of distillation under PPT-protocols, introduced by Rains \cite{rains1},
\begin{equation}
F_{d}(\rho) = \frac{I}{d} + \max {\cal f} 0, - \min_{W \in {\cal M}}Tr(W^{T_{A}}\rho) {\cal g},
\end{equation}
where ${\cal M} = {\cal f}W \hspace{0.1 cm}| \hspace{0.1 cm} (1 - d)I/d \leq W \leq I/d, \hspace{0.3 cm} 0 \leq W^{T_{A}} \leq 2I/d {\cal g}$ \cite{foot3}. 

The last is the celebrated Wootter's concurrence of two qubits, which can be written, accordingly to Verstraete \cite{verstraete1}, as
\begin{equation}
C(\rho) = \max {\cal f} 0, -\min_{A \in SL(2,C)}Tr((\ket{A}\bra{A})^{T_{B}}\rho) {\cal g},
\end{equation}
where $\ket{A}$ denotes the unnormalized state $(A \otimes I)\ket{I}$ with $\ket{I} = \sum_{i} \ket{ii}$, det$(A) = 1$.

Assuming that the set ${\cal C}$ is also convex, which is the case of all the quantities considered in this paper, except the concurrence, it is possible to apply the concept of Lagrange duality from the theory of convex optimization to the problems represented by equation (7) \cite{boyd}. Remarkably, the {\it dual} measures obtained are those related to mixing properties, such as the robustness of entanglement \cite{vidal4}, introduced by Vidal and Tarach, and the best separable approximation measure \cite{karnas}, introduced by Karnas and Lewenstein. Moreover, since in all the cases considered here there always exist a strictly feasible point, i.e, a $W \in \textbf{relint} {\cal M}$ (denoted in the convex optimization literature by Slater condition), the optimal solution of the primal and dual problems are the same, i.e., the primal and dual measures are equal \cite{boyd, ree}.

We now show that the dual representation of the generalized robustness of entanglement $R_{G}(\rho)$ \cite{steiner}, i.e., the minimal $s$ such that 
\begin{equation}
\frac{\rho + s\pi}{1 + s}
\end{equation}
is separable, where $\pi$ is any, not necessarily separable, density matrix, is given by (7) with ${\cal M} = {\cal f} W \in {\cal W} \hspace{0.1 cm} | \hspace{0.1 cm} W \leq I {\cal g}$. Following \cite{boyd}, the Lagrangian of the problem is given by
\begin{equation}
\begin{split}
L(W, g(\sigma), Z) = Tr(W\rho) + Tr(WZ)   \\
- Tr(Z)  -  \int\limits_{\sigma \in {\cal S}} g(\sigma)Tr(W\sigma)d\sigma, 
\end{split}
\end{equation}
where $Z$, $g(\sigma)$ are the Lagrange multipliers associated with the constraints $W \leq I$ and $Tr(W\sigma) \geq 0 \hspace{0.1 cm} \forall \sigma \in {\cal S}$, respectively. Note that since the definition of EW is a composition of infinite constraints, its Lagrange multiplier is a generalized function \cite{ree}. The dual problem is then
\begin{eqnarray}
\quad \mbox{minimize} \quad &  Tr(Z) \\ 
\quad \mbox{subject to} \quad & Z \geq 0  \nonumber \\ 
 & g(\sigma) \geq 0, \hspace{0.2 cm} \forall \sigma \in {\cal S} \nonumber \\
 & \rho + Z = {\displaystyle \int\limits_{\sigma \in {\cal S}}} g(\sigma)\sigma d\sigma. \nonumber
\end{eqnarray}
Since $g(\sigma) \geq 0$, the integral in the constraints above is a separable state. Conversely, any separable state $\sigma_{o}$ is obtained with the choice of $g(\sigma) = \delta(\sigma - \sigma_{o})$. It is then easily seen that the result of (15) is the generalized robustness. The dual representation of the robustness of entanglement, $R(\rho)$, has, instead of $W \leq I$, the constraint $Tr(W\sigma) \leq 1, \hspace{0.1 cm} \forall \hspace{0.1 cm} \sigma \in S$.

The best separable approximation measure $BSA(\rho)$ \cite{karnas} is the minimum $\lambda$ such that there exist a separable state $\sigma$ and a density operator $\delta \rho$ satisfying
\begin{equation}
\rho = (1 - \lambda)\sigma + \lambda \delta \rho.
\end{equation}
It can been seen that the dual representation of $BSA(\rho)$ is given by (7) with ${\cal M} = {\cal f} W \in {\cal W} \hspace{0.1 cm} | \hspace{0.1 cm} W \geq -I {\cal g}$.

In \cite{brandao2} it was shown that the random robustness $R_{r}(\rho)$ \cite{vidal4}, i.e., to the minimal $s$ such that
\begin{equation}
\frac{\rho + s(I/D)}{1 + s}
\end{equation}
is separable, is equal to equation (7), with ${\cal M} = {\cal f} W \in {\cal W} \hspace{0.1 cm} | \hspace{0.1 cm} Tr(W) = D {\cal g}$. This result can also be easily derived using the concept of Lagrange duality. 

In the next subsection we will introduce our new family of entanglement monotones

\subsection{A New Family of Entanglement Monotones}

If we let ${\cal C}$ be the set of Hermitian matrices $W$ such that $-nI \leq W \leq mI$, where $n, m \geq 0$, then the quantity derived from (7) will be denoted by $E_{n:m}$.

\begin{proposition}
$E_{n:m}$ is an entanglement monotone for every $n, m \geq 0$, i.e. 
\begin{equation}
\sum_{i}p_{i}E_{n:m}(\rho_{i}') \leq E_{n:m}(\rho)
\end{equation}
where $\rho_{i}'$ is the final state conditional on the occurrence of the classical variable $i$, which occurs with probability $p_{i}$ at the end of a LOCC protocol.
\end{proposition}

\begin{proof}
It suffices to consider final states of the form
\begin{equation}
\rho_{i}' = A_{i}\rho A_{i}^{\cal y} / p_{i}
\end{equation}
with $p_{i} = Tr[A_{i}\rho A_{i}^{\cal y}]$, where the Kraus operators $A_{1}, ..., A_{M}$ are given by $A_{i} = A_{i}^{1} \otimes ... \otimes A_{i}^{k}$ and satisfy $\sum_{i=1}^{M}A_{i}^{\cal y}A_{i} \leq I$:  
\begin{equation}
\begin{array}{c}
\sum_{i}p_{i}E_{W}(\rho_{i}') = \sum_{i}p_{i} \max{\cal f}0, -Tr(W_{\rho_{i}'}\rho_{i}') {\cal g} \\
= \sum_{k} -Tr(A_{k}^{\cal y}W_{\rho_{k}'}A_{k}\rho) \leq -Tr(W_{\rho}\rho) = E_{W}(\rho)
\end{array}
\end{equation}
where $k$ sums only the terms such that $\max{\cal f}0, -Tr(W_{\rho_{i}'}\rho_{i}'){\cal g}$ is different from zero. In the last inequality we used that $W = \sum_{k}A_{k}^{\cal y}W_{\rho_{k}'}A_{k} \leq m\sum_{k}A_{k}^{\cal y}A_{k} \leq mI$, $W = \sum_{k}A_{k}^{\cal y}W_{\rho_{k}'}A_{k} \geq -n\sum_{k}A_{k}^{\cal y}A_{k} \geq -nI$,  and that $W_{\rho}$ is optimal. 
\end{proof}
Note that the proof of proposition (2), with  minors modifications, applies also to ${\cal N}$ and $F_{d}$. 

The dual representation of $E_{m:n}(\rho)$ is
\begin{eqnarray}
\quad \mbox{minimize} \quad &  ms + nt \\ 
\quad \mbox{subject to} \quad & \rho + s\pi_{1} = (1 + s - t)\sigma + t\pi_{2} \nonumber     
\end{eqnarray}
where $\pi_{i}$ are density matrices, $\sigma$ is a separable state and $s, t \geq 0$. From (24) we find that
\begin{equation*}
\lim_{m \rightarrow \infty}E_{n:m}(\rho) = nBSA(\rho), \hspace{0.4 cm}  
\lim_{n \rightarrow \infty}E_{n:m}(\rho) = mR_{G}(\rho)
\end{equation*}
Actually, the equalities above are already valid when one of the numbers is sufficiently larger than the other. The elements of this new family of EMs can be interpreted as intermediate measures between the generalized robustness and the best separable approximation. Note that for every distinct rational number $n/m$ within a certain finite interval, the $E_{m:n}$ are genuine different EMs, meaning that there is no positive number $c$ such that $E_{m:n} = cE_{m':n'}$ if $n/m \neq n'/m'$.

If we consider that ${\cal C}$ is the intersection of set of Hermitian matrices $W$ such that $-nI \leq W \leq mI$ with the set of decomposable entanglement witnesses, a new family of entanglement monotones, denoted by $E^{PPT}_{n:m}$ is defined. To see that they are indeed EMs, all we have to note is that for every $A_{i} = A^{1}_{i} \otimes ... \otimes A^{k}_{i}$, $A^{\cal y}_{i}WA_{i}$ is a decomposable EW whenever $W$ is. Therefore, proposition (2) also applies to them. 

It is possible to derive several other families of EMs considering intersections of the sets ${\cal C}$ of different entanglement measures which can be written as equation (7), such as those given by equations (10-12). 

\section{Multipartite Entanglement Hierarchy}

We now discuss more about the different kinds of multipartite entanglement introduced in the second section. Usually the set of separable states is regarded to be composed of all sates which can be created by LOCC protocols. In this sense, given a specific split and considering that each part of the split can perform global quantum operations on its subsystems, only 1-separable states can be properly identified as separable. However, one might also be interest in the situation where some of the parties are allowed to perform join operations. In this case, the different types of entanglement play an important role. Consider, for example, the situation where $k$ parties want to create a common quantum state and each one is connected to the others via a quantum channel. If they all agree in using their channels, every state can be prepared and the situation becomes trivial. However, suppose that they agree that only $m \leq k$ parties will use their quantum channels, where the probabilities of which parties will be involved are given by $p_{i}$. At the end of the protocol they will share an ensemble of states, ${\cal f}\rho_{i}, p_{i}{\cal g}$, which clearly does not have $m + 1$-partite entanglemnt. Now, since erasing classical information cannot create entanglement, we are lead to consider the different kinds of entanglement discussed before. This property is reflected in the condition that every goog entanglement measure should be convex, which we show for every quantity defined according to equation (7)

\begin{proposition}
$E$ is a convex function for any choice of ${\cal C}$, i.e.,
\begin{equation}
E\left(\sum_{i}p_{i}\rho_{i} \right) \leq \sum_{i}p_{i}E(\rho_{i}),
\end{equation}
whenever the $\rho_{i}$ are Hermitian, and $p_{i} \geq 0$ with $\sum_{i}p_{i} = 1$.
\end{proposition}

Proposition (3) follows from the convexity and the concavity of the \textit{max} and \textit{min} functions, respectively. 

Consider a given $k$-partite split of a multi-partite system $\rho$. It is possible to attribute $(k - 1)$ numbers, $E^{m}$, $1 \leq m \leq k - 1$, where each one quantifies one type of multi-partite entanglement of the system. It is easy to see from equation (3) that all constraints imposed to an EW which detects $m$-partite entanglement ($m$-EW), are also imposed to every $n$-EW, with $n \geq m$. Hence, the following order between the $E^{m}$ holds:
\begin{equation}
E^{m}(\rho) \geq E^{n}(\rho), \hspace{0.4 cm} \forall \hspace{0.2 cm} n \geq m. 
\end{equation}

$E^{1}(\rho)$, formed by the OEW with respect to the fully separable states is an upper-bound to all other $E(\rho)$, including those with respect to other splits formed by grouping several original parties into one. This means, for example, that in a  3-split, $E^{1}(\rho)$ is greater or equal to the bipartite entanglement of any of the three 2-splits, namely A-BC, AB-C and AC-B. Actually, it is possible to establish a complete hierarchy in the proposed measures \cite{brandao3}.   

An interesting measure of entanglement for multi-partite systems is the {\it localizable entanglement}, introduced by Verstraete {\it et al} \cite{verstraete2}. Given a quantum system of $n$ parties $\rho$, the {\it localizable entanglement} $E_{ij}(\rho)$ is the maximal amount of entanglement that can be created, on average, between the parties $i$ and $j$ by performing a single-copy LOCC protocol in the system \cite{foot4}. More specifically, if at the end of a LOCC protocol we have an ensemble of states $\mu = {\cal f}p_{k}, \rho_{k}^{ij}{\cal g}$, where $p_{k}$ is the probability that the reduced state of the parties $i$ and $j$ is $\rho_{k}^{ij}$, the LE is then given by
\begin{equation}
E^{ij} = \max_{\mu}\sum_{k}p_{k}E(\rho_{k}^{ij}),
\end{equation}
where $E(\rho)$ represents, in this paper, one measure based on OEWs. The LE has the operational meaning which applies to situations in which out of some multipartite entangled state one would like to concentrate as much entanglement as possible in two particular parties \cite{verstraete2}, which could be used later, for instance, in some quantum information task. 
\begin{proposition}
Consider a multi-partite state $\rho$. Then
\begin{equation}
E^{ij}_{n:m} \leq E^{1}_{n:m}(\rho), \hspace{0.4 cm} \forall \hspace{0.2 cm} i, j, n, m.
\end{equation}
\end{proposition}

\begin{proof}
As in the proof of proposition 2, it suffices to consider final states of the form
\begin{equation}
\rho_{l}^{ij} = Tr_{/ij}(A_{l}\rho A_{l}^{\cal y} / p_{l})
\end{equation}

with $p_{l} = Tr[A_{l}\rho A_{l}^{\cal y}]$, where $Tr_{/ij}$ stands for the partial trace of all parties, except i and j. The Kraus operators $A_{1}, ..., A_{M}$ are given by $A_{l} = A_{l}^{1} \otimes ... \otimes A_{l}^{k}$ and satisfy $\sum_{i=1}^{M}A_{i}^{\cal y}A_{i} \leq I$.
\begin{equation}
\begin{array}{c}
E^{ij}_{n:m} = \sum_{l}p_{l}E_{n:m}(\rho_{l}^{ij}) = \sum_{l} \max{\cal f}0, -Tr(I\otimes W_{\rho_{l}^{ij}}\rho_{l}) {\cal g} \\
= \sum_{k} -Tr(A_{k}^{\cal y}I\otimes W_{\rho_{k}^{ij}}A_{k}\rho) \leq -Tr(W_{\rho}\rho) = E_{n:m}(\rho)
\end{array}
\end{equation}
where $k$ sums only the terms such that $\max{\cal f}0, -Tr(W_{\rho_{l}'}\rho_{l}'){\cal g}$ is different from zero. In the last inequality we used that the EW $W = \sum_{k}A_{k}^{\cal y}I\otimes W_{\rho_{k}}^{ij}A_{k} \leq m\sum_{k}A_{k}^{\cal y}A_{k} \leq mI$, $W = \sum_{k}A_{k}^{\cal y}I\otimes W_{\rho_{k}}^{ij}A_{k} \geq -n\sum_{k}A_{k}^{\cal y}A_{k} \geq -nI$ and that $W_{\rho}$ is optimal. Note that proposition (5) also applies to $E_{n:m}^{PPT}$. 
\end{proof}

The following relation between the negativity, ${\cal N}(\rho)$, and $E_{\infty:1}^{PPT}(\rho) = R_{G}^{PPT}(\rho)$ holds:
\begin{proposition}
\begin{equation}
{\cal N}(\rho) \leq E_{\infty:1}^{PPT}(\rho) \leq d{\cal N}(\rho)
\end{equation}
\end{proposition}   
\begin{proof}
For every positive operator $M$, we have
\begin{equation}
\lambda_{max}(M^{T_{A}}) \leq \lambda_{max}(M) \leq d\lambda_{max}(M^{T_{A}})
\end{equation}
where the first (second) inequality is saturated for separable (singlet) states. Hence, as $0 \leq W \leq 1$ implies $W^{T_{A}} \leq 1$, we find
\begin{eqnarray}
{\cal N}(\rho) =  -\min_{0 \leq W \leq I}Tr(W^{T_{A}}\rho) \nonumber
\\ \leq  -\min_{\substack{W^{T_{A}} \leq I \\ W \geq 0}}Tr(W^{T_{A}}\rho) = E_{\infty,1}^{PPT}(\rho)
\end{eqnarray}
where we have used that the optimal decomposable EW for a bipartite system has always the form $W^{T_{A}}$, $W \geq 0$. From equation (32) we also find that $W \geq 0$, $W^{T_{A}} \leq I$ implies $0 \leq W \leq dI$. Thus,
\begin{eqnarray}
E_{\infty:1}^{PPT}(\rho) =  -\min_{\substack{W^{T_{A}} \leq I \\ W \geq 0}}Tr(W^{T_{A}}\rho) \nonumber
\\ \leq -\min_{0 \leq W \leq dI}Tr(W^{T_{A}}\rho) = d{\cal N}(\rho)
\end{eqnarray}
\end{proof}
The second inequality is strict for example on the state
\begin{equation}
\rho = \frac{I - d(P^{+})^{T_{A}}}{d^{2} - d}
\end{equation}
where $P^{+}$ is the maximal $d$ x $d$ entangled state. 

\section{Numerical calculation}

The lack of an operational procedure to calculate entanglement measures in general is ultimately related to the complexity of distinguish entangled from separable mixed states, which was shown to be NP-HARD \cite{gurvits}. Since an operational measure, which has positive value in every entangled state, would also be a necessary and sufficient test for separability, we should not expect to find one. Nonetheless, some approximative numerical methods based on convex optimization have been proposed to the separability problem \cite{brandao1,doherty1,eisert1}. What we will show in this section is that these methods can also be used to calculate, approximately, the {\it witnessed entanglement}.         

The first one, proposed in \cite{brandao1} by Brandao and Vianna, linked the optimization of EWs with a class of convex optimization problems known as robust semidefinite programs (RSDP). Although RSDPs belong to NP-HARD, some well known probabilistic relaxations, which transforms the problem in a semidefinite program (SDP), were applied, leading to a method of optimizing {\it pseudo}-EWs (operators which are positive in almost all separable states) to every multipartite state and with respect to all types of entanglement.                

The second approach, due to Doherty {\it et al} \cite{doherty1,doherty2} , was actually the first method to the separability problem based on SDP. Using the existence of symmetric extensions for separable states and the concept of duality in convex optimization, a hierarchy of SDPs, where the $(k + 1)^{th}$ test is at least as powerful as $(k)^{th}$ (but demands more computational effort), was constructed to detect entanglement. In each step k, an OEW with respect to a restricted set of EWs, which converges to the whole set of EWs in the limit of $k \rightarrow \infty $, is obtained. This method can be used, however, only for the entanglement with respect to the fully-separable states, $E^{1}$. Note that the further constraints that we demand to the EWs can be incorporated in the SDP, since they are linear matrix (in)equalities.                     

The last method, introduced by Eisert {\it et al} \cite{eisert1}, is based on recently developed relaxations of non-convex polynomial problems of degree three in a hierarchy of SDPs, which converges to the solution of the original problem as the dimension of the SDP reaches the infinity \cite{lassere}. One of the applications of this method is the minimization of the expectation values of EWs with respect to pure product states. Therefore, it can be used together with the second method discussed to lower the value of $Tr(W\rho)$, where $W$ is a non-optimal EW determined by some step of the hierarchy.    

We would like to stress the complementary character of these methods. Whereas the first method usually provides upper bounds to $E(\rho)$, since it only determines {\it pseudo}-EWS, the second and third provides lower bounds to $E(\rho)$, as the EWs resulting from them are non-optimal. Although only the first one can calculate $E^{m}(\rho)$, for $m > 1$, the number of constrains imposed grows exponentially with $m$. Thus, in most of cases, we will restrict ourselves to the determination of $E^{1}$, which is an upper bound to all other types of multi-partite entanglement (see section IV).                                  

Note that measures restricted to decomposable EWs can always be exactly calculated, in the worst case, by a semidefinite program.           

\subsection{Example I}

\begin{figure}
\begin{center}
\includegraphics[scale=0.45]{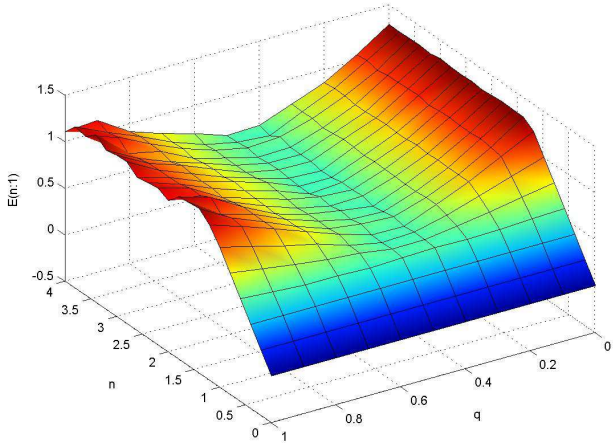}
\caption{(Color online) $E^{1}_{n:1}(\rho_{q})$ for $0 \leq n \leq 4$ and $0 \leq q \leq 1$.}
\end{center}
\end{figure}

\begin{figure}
\begin{center}
\includegraphics[scale=0.45]{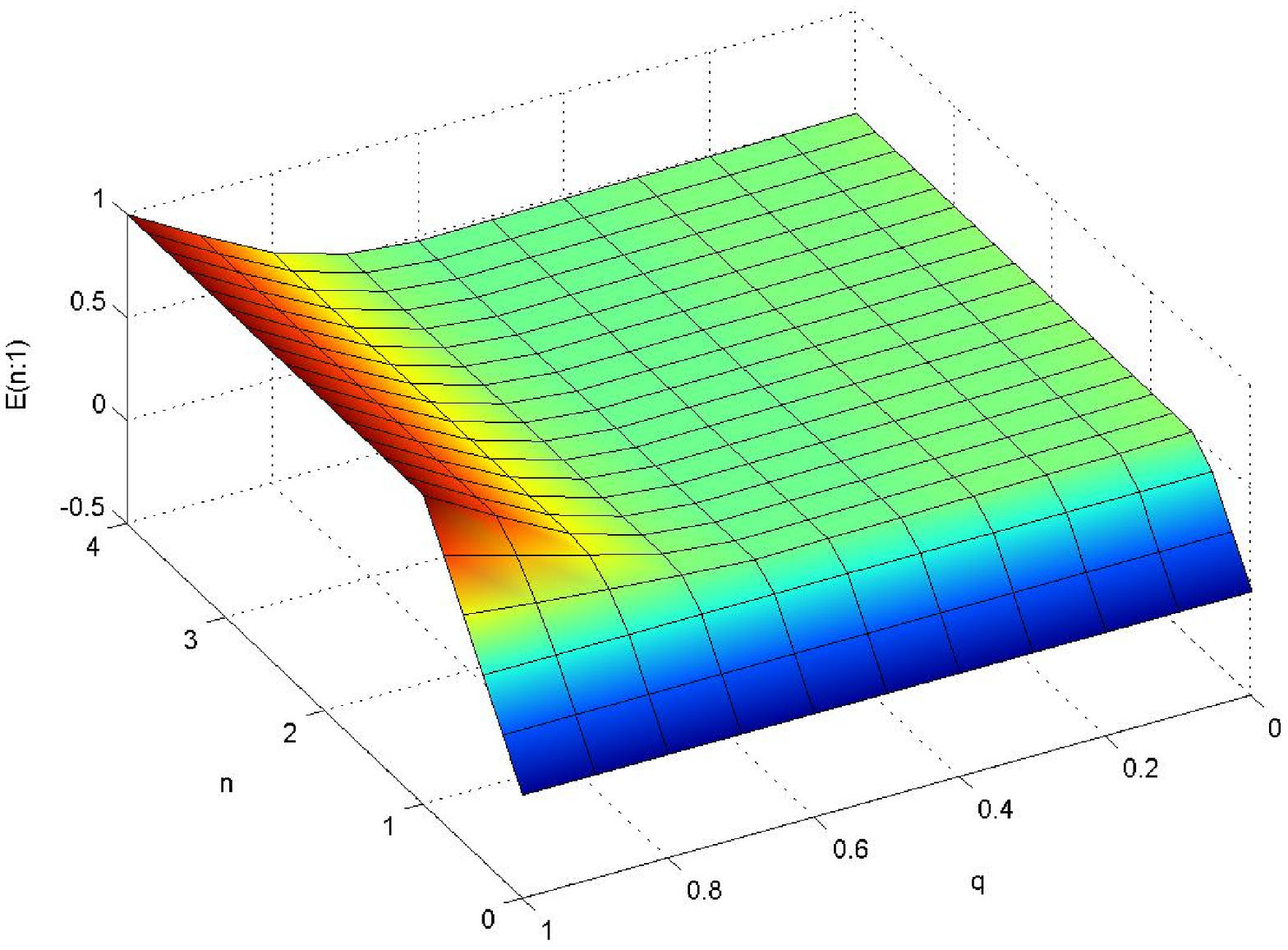}
\caption{(Color online) $E^{2}_{n:1}(\rho_{q})$ for $0 \leq n \leq 4$ and $0 \leq q \leq 1$.}
\end{center}
\end{figure}

As a first example we calculated $E_{n:1}^{1}$ and $E_{n:1}^{2}$ , $0 \leq n \leq 1$, for the following family of states
\begin{equation}
\rho_{q} = q\ket{W}\bra{W} + (1- q)\ket{GHZ}\bra{GHZ}, \hspace{0.2 cm} 0 \leq q \leq 1
\end{equation}
where $\ket{W} = (\ket{001} + \ket{010} + \ket{100})/\sqrt{3}$ and $\ket{GHZ} = (\ket{000} + \ket{111})/\sqrt{2}$. The results are plotted in figures (1) and (2). When $n << 1$, $E_{n:1}^{1} = nBSA$ and we see that for all q there is not product vectors and even biseparable vectors in the range of $\rho_{q}$. In the other limit, where $E_{n:1}^{1} = R_{G}$, we find that the generalized robustness of entanglement with respect to biseparable states is the same for all $\rho_{q}$ with $q \leq 0.7$.

\subsection{Example II}

The classes of entangled states equivalent by SLOCC for 2 X 2 X n systems were determined in \cite{miyake} and can be represented by the states (1-5) of figure (3). The arrows indicate which transformations are probabilistic possibles. 
\begin{figure}
\begin{center}
\includegraphics[scale=0.4]{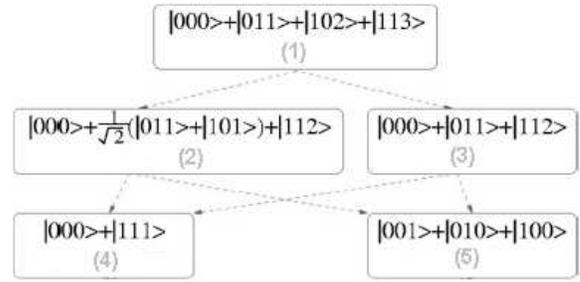}
\caption{Representative states of the five distinct classes of 3-entangled states.}
\end{center}
\end{figure}
$E_{n:1}^{1}$ , $0 \leq n \leq 1$, was calculated for each of these and plotted in figure (4).
\begin{figure}
\begin{center}
\includegraphics[scale=0.4]{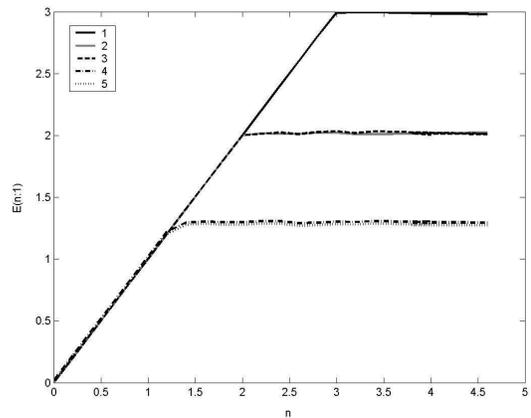}
\caption{$E^{1}_{n:1}(\rho_{q})$, $0 \leq n \leq 4.5$, for the states (1-5) of figure (3).}
\end{center}
\end{figure}
Note that for all n considered, the incomparable states (2-3) and (4-5) with respect to state transformations have approximately the same $E^{1}_{n:1}$.

\subsection{Example III}

As a final example, we present a numerical comparison between ${\cal N}$ and $E_{\infty,1} = R_{G}^{PPT}$. The bipartite PPT-generalized robustness can be determined as easily as the negativity. Actually, it can be written as
\begin{equation}
R_{G}^{PPT}(\rho) = \frac{1}{\lambda_{max}(P^{T_{A}})}{\cal N}(\rho)
\end{equation}
where $\lambda_{max}(P^{T_{A}})$ is the maximum eigenvalue of the partial transposed projector onto the negative eigenspace of $\rho^{T_{A}}$. We have generated $10^{5}$ random states using the algorithm presented in \cite{zyc} and plotted in figures (5) and (6).

\begin{figure}
\begin{center}
\includegraphics[scale=0.45]{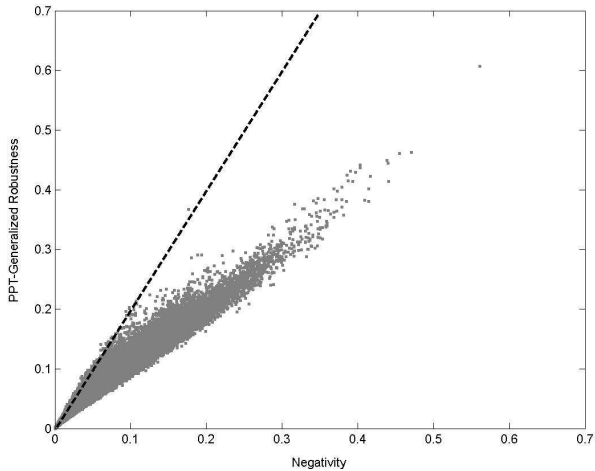}
\caption{(Color online) $R_{G}^{PPT} x \hspace{0.1 cm} {\cal N}$ for $10^{5}$ 4 x 4 random sates.}
\end{center}
\end{figure}

\begin{figure}
\begin{center}
\includegraphics[scale=0.45]{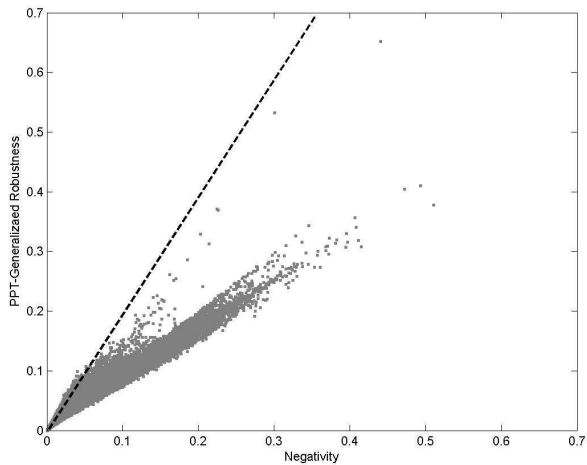}
\caption{(Color online) $R_{G}^{PPT} x \hspace{0.1 cm}  {\cal N}$ for $10^{5}$ 6 x 6 random sates.}
\end{center}
\end{figure}

Although  $d{\cal N} \geq R_{G}^{PPT} \geq {\cal N}$ (see section IV), we see from figures 5 and 6 that $R_{G}^{PPT} \leq 2{\cal N}$ for the majority of states. 
  
\section{Gaussian states}

We considered $n$ distinguishable infinite dimensional subsystems, each with local Hilbert space ${\cal H} = {\cal L}^{2}({\cal R}^{n})$. A Gaussian state is characterized by a density operator whose characteristic function $\chi_{\rho}(x) = Tr[\rho W(x)]$ is a Gaussian function \cite{giedke1}. We can write, for every Gaussian state $\rho$,
\begin{equation}  
\rho = \pi^{-n}\int_{{\cal R}^{2n}}dx e^{-\frac{1}{4}x^{T}\gamma x + id^{T}x}W(x),
\end{equation}
where $W(x) = \exp[-ix^{T}R]$ are the displacement operators and $R = (X_{1}, P_{1}, X_{2}, ..., P_{n})$, with $[X_{k}, P_{l}] = i\delta_{kl}$. The matrix $\gamma \geq iJ_{n}$ is a 2$n$ x 2$n$ real matrix called correlation matrix (CM) and $d$ is an 2$n$ real vector called displacement \cite{giedke1}. The symplectic matrix is given by
\begin{equation}
J_{n} = \bigoplus^{n}_{k = 1}J_{1}, \hspace{0.3 cm} J_{1} = \left(
\begin{array}{cc}
0 & -1 \\ 1 & 0
\end{array}
\right)
\end{equation}  
Note that the displacement of a state can always be adjusted to $d = 0$ by a sequence of unitaries applied to individual modes. This implies that d is irrelevant for the study of entanglement. Thus, we set $d = 0$ for now on without loss of generality.

The optimization of EWs for states of infinite dimension is completely infeasible. Nonetheless, we can still obtain meaningful quanties if we restrict it to a simpler, but sufficiently large, set of operators. An obvious choice would be the restriction to Gaussian entanglement witnesses (GEW), i.e., Gaussian operators which are positive in separable Gaussian states. Unfortunately, none Gaussian entangled state is detected by a GEW. Assuming that G is a GEW with covariance matrix $\Gamma$, we find 
\begin{equation}
Tr(\rho G) = \int_{{\cal R}^{2n}}dx e^{-\frac{1}{4}x^{T}(\Gamma + \gamma)x + c} \geq 0. 
\end{equation}

Another possible class of operators then is given by
\begin{equation}
{\cal W}_{\cal G} = {\cal f} Q \in {\cal B}({\cal H} \otimes {\cal H}) \hspace{0.1 cm} | \hspace{0.1 cm} Q = 2^{n}I - G {\cal g},
\end{equation}
where $G$ is a Gaussian operator and $I$ the identity operator \cite{foot5}. In the next proposition we show that $E_{\infty:m}^{G}(\rho)$ given by
\begin{equation}
E_{\infty:1}^{G}(\rho) = \max {\cal f} 0, -\min_{Q \in {\cal W}_{\cal G}, \hspace{0.1 cm} Q \leq I} Tr(Q \rho) {\cal g}
\end{equation}
can be very efficiently numerically calculated by a simple semidefinite program.

\begin{proposition}

\begin{equation}
E_{\infty:m}^{G}(\rho) = \max {\cal f}0, \hspace{0.1 cm} \det(\Gamma + \gamma)^{-1/2} - 2^{n} {\cal g}
\end{equation}
where the matrix $\Gamma \in M_{2n}({\cal R})$ is obtained by the following SDP determinant maximization problem
\begin{equation*}
\max_{\Gamma, \tau, \Delta, s_{ij}} \hspace{0.1 cm} \det(\Gamma + \gamma)^{-1/2}
\end{equation*}
\begin{center}
subject to  \hspace{0.3 cm} $-1 \leq \tau \leq 1, \hspace{0.6 cm} \Gamma \geq 0$ 
\end{center}
\begin{eqnarray}
\left( 
\begin{array}{cc} 
\tilde{\Gamma}_{2} + i\tau J & \tilde{\Gamma}_{12}^{T} \\
\tilde{\Gamma}_{12} & \tilde{\Gamma}_{1} - iJ
\end{array} \right) \geq 0 
\end{eqnarray}

\begin{equation*}
\left(
\begin{array}{cc}
J_{n} + \Gamma & \Delta \\
\Delta & D(\Delta)
\end{array} \right) \geq 0, \hspace{0.4 cm} \left(
\begin{array}{cc}
s_{k-1,2l-1} & s_{k,l} \\
s_{k,l} & s_{k-1,2l}
\end{array} \right) \geq 0 
\end{equation*}
\begin{center}
$s_{k-1,2l-1} \geq 0, \hspace{0.11cm} s_{k-1,2l} \geq 0, \hspace{0.11cm} s_{kl} \geq 2$, \hspace{0.11 cm}   {\small $k = 1..l, i = 1...2^{l-k}$}
\end{center}
where $\Delta$ is a n x n lower triangular matrix comprised of additional variables, D($\Delta$) is a diagonal matrix with same diagonal entries as those of $\Delta$, l is the smallest number such that $2^{l} \geq n$, and $s_{0,i} = \Delta_{ii}$ if $1 \leq i \leq n$ and $s_{0,i} = 2$ if $n \leq i \leq 2^{l}$. 
 
\end{proposition}

\begin{proof}

Consider the following structure for the bipartite Gaussian operator $G$ 
\begin{equation}  
G = \int_{{\cal R}^{2n}}dx e^{-\frac{1}{4}x^{T}\Gamma x}W(x),
\end{equation}
where $\Gamma^{T} = \Gamma \geq 0 \in M_{2n}({\cal R})$, with modes $1$ to $m$ and $m + 1$ to $n$ belonging to Alice and Bob, respectively. The optimization objective $Tr(Q \rho)$, where $Q = 2^{n}I - G$, can be written as
\begin{equation}
2^{n} - \pi^{-n}\int_{{\cal R}^{2n}}dx e^{-\frac{1}{4}x^{T}(\Gamma + \gamma)x}
= 2^{n} - \det(\Gamma + \gamma)^{-1/2}
\end{equation}

From the Jamiolkowski isomorphism, $Q$ is an EW iff the map ${\cal Q}$ defined as $Q = I \otimes {\cal Q}(P^{+}) = 2^{n}I - I \otimes {\cal G}(P^{+})$ \cite{foot6} is positive, which is equivalent to $\rho' = I \otimes {\cal G}(\rho) \leq 2^{n}I$, for every density operator $\rho$. The covariance matrix of $\rho'$, $\gamma'$, can be written as \cite{simon} 
\begin{equation}
\gamma' = S^{T}\sigma S ,
\end{equation}
where $S \in Sp(2n,{\cal R})$  and $\sigma$ is the covariance matrix
\begin{equation*}
\sigma = \mbox{diag}(\mu_{1},\mu_{1},...,\mu_{n},\mu_{n})
\end{equation*}
corresponding to a tensor product of states diagonal in the number basis given by
\begin{equation}
M' = \bigotimes_{i}\frac{2}{\mu_{i} + 1}\sum_{k=0}^{\infty}\left(\frac{\mu_{i} - 1}{\mu_{i} + 1}\right)\ket{k}_{i}{}_{i}\bra{k},
\end{equation} 
$\ket{k}_{i}$ being the {\it k}-th number state of the Fock space ${\cal H}_{i}$ \cite{adesso}. The symplectic transformation (47) is reflected in the Hilbert space level by an unitary transformation: $G = U(S)^{\cal y}G'U(S)$. Since we are considering bounded operators, the $\mu_{i}$ must be non-negative. We also see that the positiveness of ${\cal Q}$ is equivalent to 
\begin{equation}
\lambda_{max}(\rho') = \prod_{j = 1}^{n} \left( \frac{1}{1 + \mu_{j}} \right) \leq 1, \hspace{0.2 cm} \forall \rho' 
\end{equation}
Thus, since we are only considering Gaussian operators, equation (49) is satisfied iff $\gamma' \geq 0$ for every $\gamma \geq iJ$, where $\gamma'$ and $\gamma$ are the covariance matrices of $\rho'$ and $\rho$, respectively. Following Giedke and Cirac \cite{giedke1}, one finds that ${\cal Q}$ is positive iff
\begin{equation}
\min_{z \in{\cal C}^{2n}}\max{\cal f}z^{\cal y}(M + iJ)z, z^{\cal y}(M - iJ)z{\cal g} \geq 0
\end{equation}         
where $M = \tilde{\Gamma}_{2} - \tilde{\Gamma}_{12}^{T}(\tilde{\Gamma}_{1})^{-1}\tilde{\Gamma}_{12}$ and $\tilde{\Gamma} = (I \oplus \Lambda)\Gamma(I \oplus \Lambda)$, with $\Lambda = $ diag$(1,-1,1,-1,...,-1)$. The matrices $\Gamma_{i}$ are such that
\begin{equation*}
\Gamma = \left(
\begin{array}{cc}
\Gamma_{1} & \Gamma_{12} \\
\Gamma_{12}^{T} & \Gamma_{2}
\end{array} \right)
\end{equation*}

We now express condition (50) as a linear matrix inequality. Equation (50) is equivalent to
\begin{eqnarray}
z^{\cal y}(M + iJ)z \geq 0 \quad \mbox{$\forall \hspace{0.1 cm} z \in {\cal C}^{2n}$ \hspace{0.1 cm}s.t.} \quad z^{\cal y}(M - iJ)z \leq 0 \nonumber \\
z^{\cal y}(M - iJ)z \geq 0 \quad \mbox{$ \forall \hspace{0.1 cm} z \in {\cal C}^{2n}$ \hspace{0.1 cm} s.t.} \quad z^{\cal y}(M + iJ)z \leq 0 \nonumber
\end{eqnarray}

An important theorem of matrix analysis, known as {\cal S}-procedure, can be stated as follows: a quadratic function in the variable $x$, $G(x)$, is positive for all $x$ such that $H(x) \geq 0$, where $H(x)$ is another quadratic function, iff there exists a positive real number $\tau$ such that $G - \tau H \geq 0$, for all $x$ \cite{boyd2}.    

Applying it to the two conditions above we find that equation (50) holds iff there exists a positive number $\tau$ such that
\begin{equation}
M + \tau iJ \geq 0, \hspace{0.3 cm} -1 \leq \tau \leq 1
\end{equation}         
We now use another fact of matrix analysis which says that the constraints on the Schur complement $R > 0, \hspace{0.3 cm} Q - SR^{-1}S^{T} \geq 0$ and $\ker(R) \subseteq \ker(Q)$ are equivalent to
\begin{equation*}
\left( \begin{array}{cc}
Q & S \\ S^{T} & R
\end{array} \right) \geq 0
\end{equation*}  
Hence, applying it to equation (51), we find that a Gaussian operator ${\cal G}$ is an EW iff there exists a real number $-1 \leq \tau \leq 1$ such that equation (44) holds. 

From the Williason decomposition, we see that $Q \leq I$ is equivalent to
\begin{equation} 
\prod_{i=1}^{n}\left(\frac{2}{\eta_{i} + 1}\right) = 2^{n}\det(I + (S^{T})^{-1}\Gamma S^{-1})^{-1} \leq 1,
\end{equation}
where $\eta_{i}$ ate the symplectic eigenvalues of $Q$. Since $S$ is symplectic, one has $S^{T}J_{n}S = J_{n}$, so that $\det(I + (S^{T})^{-1}\Gamma S^{-1})^{-1} = \det[{J_{n}^{T}(S^{T})^{-1}(J_{n} + \Gamma)S^{-1}}]^{-1} = \det(J_{n} + \Gamma)^{-1}$ . The proposition then follows from reference \cite{ben}, which presents a LMI representation for the inequality $\det(A)^{1/m} \geq t$, where $A$ is a positive $m$ x $m$ real matrix.  
\end{proof}

\section{States with symmetry}

Entanglement measures have usually their calculation greatly simplified when the state in question has certain symmetries. Following \cite{vollbrecht1}, let $G$ be a closed group of product unitary operators of the form $U = U_{1} \otimes U_{2}$. Defining the projection
\begin{equation}
\textbf{P}(A) = \int dU \hspace{0.2 cm} UAU^{*},
\end{equation}
for any operator $A$ on $H_{1}\otimes H_{2}$ \cite{foot7}, where $dU$ is the Haar measure of $G$, we say that an operator $M$ is invariant under $G$ if $\textbf{P}(M) = M$, which is equivalent to $[U, M] = 0$ for all $U \in G$. Consider now the determination of any measure expressed by equation (7). If the state in question has the property $\textbf{P}(\rho) = \rho$, then one can restrict the optimization in (7) to operators with the this same symmetry. More specifically,
\begin{eqnarray}
E(\rho) = -\min_{W \in {\cal M}} Tr(W\rho) = -\min_{W \in {\cal M}} Tr(W\textbf{P}(\rho))  \nonumber
\\ = -\min_{W \in {\cal M}} Tr(\textbf{P}(W)\rho) = -\min_{W \in \textbf{P}({\cal M})} Tr(W\rho), 
\end{eqnarray} 
where  $\textbf{P}({\cal M}) = {\cal f}W \in {\cal M} \hspace{0.1 cm} | \hspace{0.1 cm} W = \textbf{P}(W){\cal g}$ \cite{foot10}. 

As an example, consider the isotropic states $\rho_{p}$ on ${\cal C}^{d}\otimes{\cal C}^{d}$
\begin{equation}
\rho_{p} = pP^{+} + (1-p)\frac{I}{d^{2}},
\end{equation}
where $P^{+} = \ket{\Phi^{+}}\bra{\Phi^{+}}$ is the maximally entangled state. It can be shown that $P^{+}$ and the identity are the only operators which commute with all unitaries of the form $U\otimes U^{*}$. Hence, the OEWs for $\rho_{p}$ can be written as
\begin{equation}
W(\rho_{p}) = a(p)P^{+} + b(p)I.
\end{equation}
Since $\bra{\psi}P^{+}\ket{\psi} \leq 1/d$, for every separable state $\ket{\psi}$, we find
\begin{eqnarray}
E_{n:1}(\rho_{p}) = \begin{cases}
(n + 1)p + \frac{(1 - p)(n + 1)}{d^{2}} - 1 &  n \leq d-1, \\
dp + \frac{1 - p}{d} - 1 &  n \geq d.
\end{cases}
\end{eqnarray}
As the OEWs for this family of states are decomposable, equation (57) is also valid to $E^{PPT}_{n:1}$.

\section{Superselection Rules}

The effect of superselection rules (SSR) in theory of entanglement has been studied recently under a number of different perspectives \cite{verstraete3,bartlett1,terhal3,schuch}. Two striking features emerge from the existence of a SSR. The entanglement of a given state under SSR is usually reduced \cite{bartlett1} and the notion of nonlocality has to be redefined, as there exists separable states that cannot be created by LOCC \cite{verstraete3}. In this section we show how the {\it witnessed entanglemet} fit in each of these scenarios.   

Following Bartlett and Wiseman \cite{bartlett1}, we define a SSR as a restriction on the allowed local operations on a system, associated with a group of physical transformations $G$. An operation ${\cal O}$ is $G$-covariant if
\begin{equation}
{\cal O}[T(g)\rho T^{\cal y}(g)] = T(g){\cal O}[\rho]T^{\cal y}(g),
\end{equation}
for all group elements $g \in G$ and all density operators $\rho$. Then the SSR associated to $G$ is the restriction on the allowed operations on the system to those $G$-invariant. As these restrictions make a state $\rho$ indistinguishable from the states $T(g)\rho T^{\cal y}(\rho)$ for all $g \in G$, it is convenient to describe $\rho$ by the $G$-invariant state
\begin{equation}
{\cal G}(\rho) = \int\limits_{G}dgT(g)\rho T^{\cal y}(g),
\end{equation}
where $dg$ is the group-invariant Haar measure. For multipartite systems, where the SSRs are local, we have ${\cal G}[\rho] = {\cal G}\otimes ... \otimes{\cal G}[\rho]$. As it was shown in \cite{bartlett1}, the maximal amount of entanglement which can be produced by LOCC in a register shared by all the parties, initially in a product state and not subjected to SSRs, from a state $\rho$, constrained by a $G$-SSR, is given by the entanglement they can produce from ${\cal G}(\rho)$ by unconstrained LOCC. If $E$ is an entanglement monotone, any LOCC applied to ${\cal G}(\rho)$, can, on average, at most maintain $E({\cal G}[\rho])$. Since it is always possible to reach this bound applying local swap operators, we have that the maximal amount of entanglement produced by SSR is exactly $E({\cal G}[\rho])$. Hence, from section (VII), it follows that, under a $G$-SSR
\begin{equation}
E(\rho) = \max {\cal f}0, -\min_{W \in {\cal G}[{\cal M}]} Tr(W\rho) {\cal g},
\end{equation}       
where  ${\cal G}[{\cal M}] = {\cal f}W \in {\cal M} \hspace{0.1 cm} | \hspace{0.1 cm} W = {\cal G}[W]{\cal g}$ \cite{foot8}.

We now consider the effect of SSRs in the notion of locality. The states that can be prepared locally in the presence of a $G$-SSR are those which can be written as (1), with each $\ket{\psi_{i}}_{k}$ being $G$-invariant. It is possible to detect nonlocal states with witness operators, defining a $G$-nonlocality witness (GW) as a Hermitian operator which satisfies equation (3), with $\ket{\psi}_{P_{k}^{m}}{}_{P_{k}^{m}}\bra{\psi} = {\cal G}[\ket{\psi}_{P_{k}^{m}}{}_{P_{k}^{m}}\bra{\psi}]$, for all $i$ and $k$. This nonlocal character of some states in the presence of a SSR can be quantified \cite{schuch}. We can then, as it was done with entanglement, use GWs to perform this quantification. A {\it witnessed nonlocality measure}, $N_{G}$, will be any quantie given by equation (7), with the set of EWs substituted by the set of nonlocality witnesses. It is easy to see that all properties discussed for $E$ are valid for $N_{G}$.  

As an example, considered the following state
\begin{eqnarray}
\rho = \frac{1}{4}(\ket{0}_{A}\bra{0}\otimes \ket{0}_{B}\bra{0} + \ket{1}_{A}\bra{1}\otimes \ket{1}_{B}\bra{1}) \nonumber \\  + \frac{1}{2}\ket{\Psi_{+}}_{AB}\bra{\Psi_{+}},
\end{eqnarray}
where $\ket{\Psi_{+}}_{AB} = (\ket{0}_{A}\ket{1}_{B} + \ket{1}_{A}\ket{1}_{B})/\sqrt(2)$. Verstraete and Cirac have shown \cite{verstraete3} that, although this state has a separable decomposition, it is not local when a particle number SSR exists, since all possible separable decompositions have local states involving superpositions of a different number of particles. Any Hermitian matrix with positive diagonal entries is a G-nonlocality witness in this case. This should be contrasted with the case of a general EW, where an infinite number of inequalities are necessary for its characterization. Calculating, for example, the nonlocal measure equivalent to $E_{\infty:1} = R_{G}$,
\begin{equation*}
N_{G}(\rho) = \max{\cal f}0, -\min_{G \in {\cal G}}Tr(G\rho){\cal g},  
\end{equation*}
where ${\cal G} = {\cal f}G \hspace{0.1 cm} | \hspace{0.1 cm} G_{ii} \geq 0, G \leq I {\cal g}$, we find $N_{G})(\rho) = 1/2$, with $G = -\ket{01}\bra{10} - \ket{10}\bra{01}$.  

\section{Indistinguishable particles}

The study of entanglement in systems of indistinguishable particles has been the subject of recent controversy \cite{zanardi1,gittings,pask,schliemann,wiseman1}. At least three different approaches to the problems have been proposed, namely, the {\it entanglement of modes} \cite{zanardi1}, the {\it quantum correlations} \cite{schliemann} and the {\it entanglement of particles} \cite{wiseman1}. Each of these has its own advantages and drawbacks, and no consensus has been reached on which one is the most suitable. In this section we show how the proposed measures based on EW can be used to quantify entanglement in each of the three methods.
   
We start with the {\it entanglement of modes}, proposed by Zanardi \cite{zanardi1}, which suggests that the entanglement of indistinguishable particles should be calculated by any regular entanglement measure, using the density matrix in the mode-occupation, or Fock, representation. In this case it is clear that the determination of the witnessed entanglement follows straightforwardly.
 
The {\it quantum correlations}, introduced by Schliemann {\it et al} \cite{schliemann}, is motivated by the believe that no quantum correlations due to symmetrization (for bosons) or anti-symmetrization (for fermions) should be considered as true entanglement. Then, the characterization of entanglement, for pure states, is determined by the Slater rank of the state, as opposed to the Schimdt rank usually considered in distinguishable particles. Furthermore, Schliemann {\it et al} have shown that the concept of entanglement witness is also applicable to multipartite systems of indistinguishable particles \cite{foot11} and, thus, the witnessed entanglement are well defined in this case too.
   
The last approach, due to Wiseman and Vaccaro \cite{wiseman1}, is probably the best motivated one. The {\it entanglement of particles} is defined as the maximal amount of entanglement, computed by a standard measure, which Alice and Bob can produce between a quantum register, shared by them, composed of distinguishable particles by local operations. The amount of entanglement will clearly depends on the physical constraints imposed, which are in most of cases expressed as a SSR. Therefore the approach presented in the previous section to SSR can also be applied in this case.            

\section{Teleportation distance} 

In this section we derive lower bounds to the teleportation distance, using $E_{n:m}$. We consider a quantum state $\rho$ shared by $k$ parties and ask what is the best possible teleportation distance attained by a LOCC protocol when the parties form two groups and teleport a quantum state from one group to the other. 

Consider a teleportation protocol where a bipartite state $\rho_{AB}$ is used as a quantum channel between Alice and Bob. Following the approach of Vidal and Werner to the negativity \cite{vidal3}, we will first consider the {\it single distance} of a bipartite state defined as:
\begin{equation}
\Delta(P_{+}, \rho) \equiv \inf_{P}||P_{+} - P(\rho)||_{1},
\end{equation}
where $P_{+}$ is the maximally entangled state and the infimum is taken over LOCC protocols $P$. Using the convexity and the invariance under unitary transformations in the two terms of the absolute distance, and the invariance of $P_{+}$ under unitary transformations of the form $U\otimes U^{*}$, we may assume that the optimal state which minimizes equation (62), $P_{opt}(\rho)$, has undergone a twirling operation \cite{foot9} and, therefore, is a {\it noise singlet},
\begin{equation}
\rho_{p} = pP_{+} + (1-p)\frac{I\otimes I}{d^{2}}.
\end{equation}
The absolute distance of $\rho_{p}$ is given by $||P_{+} - \rho_{p}||_{1} = 2(1 - p)(d^{2} - 1)/d^{2}$. From equation (57), 
\begin{equation}
||P_{+} - \rho_{p}||_{1} = 2(1 - \frac{1 + E_{n:1}^{PPT}(\rho_{p})}{d}).
\end{equation}
From the monotonicity of $E_{n:1}^{PPT}$ under LOCC, we find that, for $n \geq d$,  
\begin{proposition}
\begin{equation}
\Delta(P_{+}, \rho) \geq 2(1 - \frac{1 + E_{n:1}^{PPT}(\rho)}{d}).
\end{equation}
\end{proposition}
Since $E_{n:1}^{PPT}(\rho_p) = 2{\cal N}(\rho_{p})$, for $n \geq d$, we see that $E_{n:1}^{PPT}$ provides, when $E_{n:1}^{PPT} \leq 2{\cal N}$, a sharper bound than the one derived from the negativity. In the limit case $n \rightarrow \infty$ already, where $E_{n:1}^{PPT}$ is equal to the PPT-generalized robustness, we see from section (V) that the new bound is indeed sharper for the majority of states.

A measure of the degree of performance of a quantum channel is the teleportation distance
\begin{equation}
d(\Lambda) = \int d\phi ||\phi - \Lambda(\phi)||_{1}.
\end{equation}
As it was shown by the Horedecki family \cite{horodecki1}, the minimal teleportation distance that can be achieved when using the bipartite state $\rho$ to construct an arbitrary teleportation channel is given by
\begin{equation}
d_{min}(\rho) = \frac{d}{d + 1}\Delta(P_{+}, \rho).
\end{equation}
Therefore,
\begin{equation}
d_{min}(\rho) \leq \frac{2d}{d + 1}(1 - \frac{1 + E_{n:1}^{PPT}(\rho)}{d}), \hspace{0.1 cm} n \geq d.
\end{equation}
Until now we have just adapted Vidal and Werner's reasoning for the negativity to the $E_{n:1}^{PPT}$. Nevertheless, as opposed to ${\cal N}$, $E_{n:1}^{PPT}$ are also defined to multi-partite systems. 
\begin{proposition}
Consider a quantum state $\rho$ shared by $k$ parties. Let $\rho^{1..m:(m+1)..k}$ denote a bipartite split of the system, where the parties 1 to $m$ and $m$ + 1 to $k$ form two groups. Then,
\begin{equation}
d_{min}(\rho^{1..m:(m+1)..k}) \leq \frac{2D}{D + 1}(1 - \frac{1 + (E_{n:1}^{PPT})^{1}(\rho)}{D}),
\end{equation}
$\forall \hspace{0.1 cm} 1 < m < k$, where $D$ stands for the minimum of the dimensions of the two groups.
\end{proposition} 
  
Proposition (8) follows from the upper bound to all types of entanglement provided by $E^{1}$. Equation (57) is saturated, for example, on the $k$-partite GHZ state $\ket{\Psi_{GHZ}} = 1/\sqrt{2}(\ket{00...0} + \ket{11...1})$.          

\section{Upper Bounds for the Distillable Entanglement}

We now move on to show another application of the family $E_{1:m}$, namely bounds to the distillable entanglement of bipartite mixed states. We first derive the following additivity property

\begin{proposition}
\begin{equation}
E_{n:1}(\rho \otimes \rho) \leq E_{n:1}(\rho)^{2} + 2E_{n:1}(\rho) , \hspace{0.2 cm} \forall \hspace{0.1 cm} n \geq 1.
\end{equation}
\end{proposition}
\begin{proof}
Consider the dual representation (24) of $E_{n:1}$. Let s, t, $\sigma$, $\pi_{1}$ and $\pi_{2}$ be variables which minimize (24). Then we find $E_{n:1}(\rho) = s + nt$, with
\begin{equation}
\rho =  (1 + s - t)\sigma + t\pi_{2} - s\pi_{2}.
\end{equation}
Thus,
\begin{eqnarray}
\rho \otimes \rho = (1 + s - t)^{2}\sigma \otimes \sigma + t(1 + s - t)\sigma \otimes \pi_{2} \nonumber \\ 
- s(1 + s - t) \sigma \otimes \pi_{1} + t(1 + s - t)\pi_{2} \otimes \sigma + t^{2}\pi_{2} \otimes \pi_{2} \nonumber \\
- st\pi_{2}\otimes \pi_{1} - s(1 + s - t)\pi_{1}\otimes \sigma - st\pi_{1}\otimes \pi_{2} + s^{2}\pi_{1}\otimes \pi_{1} \nonumber  \\
= [1 + (2s + s^{2} + t^{2}) - (2t + 2st)]\sigma \otimes \sigma \nonumber \\
+ [t(\rho \otimes \pi_{2} + \pi_{2} \otimes \rho) + st(\pi_{1}\otimes \pi_{2} + \pi_{2}\otimes \pi_{1})] \nonumber \\
- [ s(\rho \otimes \pi_{1} + \pi_{1} \otimes \rho) + s^{2} \pi_{1}\otimes \pi_{1} + t^{2}\pi_{2} \otimes \pi_{2}], \nonumber
\end{eqnarray}
where in the last two lines we used that 
\begin{equation*}
\sigma = \frac{1}{1 + s - t}\left(\rho + s\pi_{1} - t\pi_{2}\right).
\end{equation*}
It is therefore easily seen that if $E_{n:1}(\rho \otimes \rho) = s' + nt'$, then $s' + nt'\leq s^{2} + t^{2} + 2s + n(2t + 2st)$. Hence, as $n \geq 1$,
\begin{eqnarray}   
E_{n:1}(\rho)^{2} + 2E_{n:1}(\rho) - E_{n:1}(\rho \otimes \rho) \nonumber 
\\ = s^{2} + n^{2}t^{2} + 2nst + 2s + 2nt - s' - nt' \nonumber \\ 
\geq s^{2} + n^{2}t^{2} + 2nst + 2s + 2nt - s^{2} - t^{2} - 2s - 2nt - 2nst \nonumber \\
= t^{2}(n^{2} - 1) \geq 0.   \nonumber
\end{eqnarray}
\end{proof}
We can define a family of quantities close related to $E_{n:1}$ by
\begin{equation}
LE_{n:1}(\rho) = \log_{2}(1 + E_{n:1}(\rho)).
\end{equation}
The $LE_{n:1}(\rho)$ are non-increasing under trace preserving separable operations. From proposition (9) we find that they are also weakly-subadditive. Indeed, for $n \geq 1$, 
\begin{equation} 
LE_{n:1}(\rho \otimes \rho) \leq \log_{2}((1 + E_{n:1}(\rho))^{2}) = 2LE_{n:1}(\rho).
\end{equation}
Note that the same results are also valid to $E^{PPT}_{n:1}$. We now can state the main result of this section
\begin{proposition}
\begin{equation}
E_{D}(\rho) \leq LE_{n:1}(\rho) , \hspace{0.1 cm} \forall \hspace{0.1 cm} n \geq 1.
\end{equation}
where $E_{D}(\rho)$ is the distillable entanglement of the bipartite state $\rho$.
\end{proposition}
\begin{proof}
The proof of proposition (10) is basically an application of a theorem due to the Horodeckis \cite{horodecki5} which can be stated as follows: any function B satisfying the conditions 1)-3) below is an upper bound for the entanglement of distillation. 
\begin{enumerate}
	\item Weak monotonicity: $B(\rho) \geq B(\Lambda(\rho))$ where $\Lambda$ is any trace-preserving superoperator realizable by means of LOCC operations.
	\item Partial subadditivity: $B(\rho^{n}) \leq nB(\rho)$.
	\item Continuity for isotropic states $\rho_{p}$ given by equation (55). Suppose that we have a sequence of isotropic states $\rho_{p}$\ such that $Tr(\rho_{p}P^{+}) \rightarrow 1$, if $d \rightarrow \infty$. Then we require
	\begin{equation}
	\lim_{d \rightarrow \infty}\frac{1}{\log_{2}d}B(\rho_{p}) \rightarrow 1
	\end{equation}	
\end{enumerate}

We have already shown that $LE_{n:1}(\rho)$, for $n \geq 1$, satisfies conditions (1) and (2). From equation (54)
\begin{equation}
LE_{n:1}(\rho_{p}) = \log_{2}(dp + \frac{1 - p}{d}), \hspace{0.1 cm} \forall \hspace{0.1 cm} n \geq 1
\end{equation}
By evaluating this expression now for large $d$, we easily obtain that condition (3) is satisfied.
\end{proof}

It is also possible to state a proposition like (8) to the bounds on the distillable entanglement. $E^{1}$ will in this case provide an upper bound to $E_{D}$ of all bipartite partitions. 

\section{Lower bounds for the entanglement of formation}

One of the most celebrated entanglement measures is the entanglement of formation \cite{bennett2}
\begin{equation}
E_{F}(\rho) = \min_{p_{1}, \psi_{i}}\sum_{i}p_{i}E_{E}(\ket{\psi_{i}}),
\end{equation}   
where $E_{E}$ is the entropy of entanglement. Although this measure has a very meaningful physical interpretation and good properties, its calculation has been done only for a very class of states \cite{terhal5}. We show in this section that any entanglement witness can be used to provide lower bounds to the entanglement of formation.

Let $\rho = \ket{\Psi}\bra{\Psi}$ be a pure bipartite state with the following Schimdt decomposition:
\begin{equation}
\ket{\Psi} = \sum_{j=1}^{d} c_{j} \ket{jj} , \hspace{0.4 cm} c_{1} \geq c_{2} \geq ... \geq c_{d}.
\end{equation}
An analytic expression for the random robustness $R_{r}$ and the generalized robustness $R_{G}$ of a pure state given by equation (78) is \cite{vidal4}
\begin{equation}
R_{G}(\rho) = \left(\sum_{j=1}^{d}c_{j} \right)^{2} - 1,
\end{equation}
\begin{equation}
R_{r}(\rho) = c_{1}c_{2}.
\end{equation} 

We start with two bounds for the entropy of entanglement, i.e., the von Neumann entropy of the reduced density matrix of a pure state $\ket{\psi}$. In the case of two qubits, Wootters has shown that \cite{wootters1} 
\begin{equation*}
H\left( \frac{1 + \sqrt{1 - 4c_{1}^{2}c_{2}^{2}}}{2}\right) = E_{E}(\ket{\psi}),
\end{equation*} 
where $H(x) = -x\log(x) - (1 - x)\log(1 - x)$. That is a particular case of the more general inequality 
\begin{equation}
H\left(\frac{1 + \sqrt{1 - 4c_{1}^{2}c_{2}^{2}}}{2}\right) \leq -\sum_{i}^{d}c_{i}^{2}\log(c_{i}^{2}), \hspace{0.2 cm} \sum_{i}^{d}c_{i}^{2} = 1.
\end{equation}  
Another similar inequality is
\begin{equation}
\frac{\log(d) - 1}{d}\left[\left(\sum_{i}^{d}c_{i}\right)- 1\right]  \leq -\sum_{i}^{d}c_{i}^{2}\log(c_{i}^{2}).
\end{equation}  
Equations (72-73) can be proved maximizing the L.H.S. minus the R.H.S. and noting that the maximum is null in both cases.   

Choosing ${\cal f}p_{i}, \ket{\psi_{i}}{\cal g}$ to be an optimal ensemble in equation (71), we have
\begin{eqnarray}
E_{F}(\rho) = \sum_{i}p_{i}E_{E}(\ket{\psi_{i}}) \geq \sum_{i}p_{i}H\left(\frac{1 + \sqrt{1 - 4(c_{1}^{2})_{i}(c_{2}^{2})_{i}}}{2}\right) \nonumber \\
\geq H\left(\frac{1 + \sqrt{1 - 4R_{r}^{2}(\rho)}}{2}\right),  \nonumber
\end{eqnarray}
where we have used the convexity of $R_{r}$ and $f(x) = H\left(\frac{1 + \sqrt{1 - 4x^{2}}}{2}\right)$. Similarly, we find
\begin{equation}
E_{F}(\rho) \geq \frac{\log(d) - 1}{d}R_{G}(\rho).
\end{equation}
The bound derived from $R_{r}$ is suitable for slightly entangled states, where the Schimdt coefficients of the optimal $\ket{\psi_{i}}$ decay fast enough, making the truncation in the second Schimdt coefficient a good approximation. 

As an first example, we consider the Horodecki 3x3 states \cite{horodecki6}. These states exhibit bound entanglement, since they have positive partial transposition. They are given by
\begin{equation}
\rho(a) = \left[
\begin{array}{ccccccccc}
a & 0 & 0 & 0 & a & 0 & 0 & 0 & a \\
0 & 0 & a & 0 & 0 & 0 & 0 & 0 & 0 \\
0 & 0 & 0 & a & 0 & 0 & 0 & 0 & 0 \\
0 & 0 & 0 & 0 & a & 0 & 0 & 0 & 0 \\
a & 0 & 0 & 0 & a & 0 & 0 & 0 & a \\
0 & 0 & 0 & 0 & 0 & a & 0 & 0 & 0 \\
0 & 0 & 0 & 0 & 0 & 0 & \frac{1 + a}{2} & 0 & \frac{\sqrt{1 - a^{2}}}{2}\\
0 & 0 & 0 & 0 & 0 & 0 & 0 & a & 0 \\
0 & 0 & 0 & 0 & 0 & 0 & \frac{\sqrt{1 - a^{2}}}{2} & 0 & \frac{1 + a}{2}
\end{array}
\right]
\end{equation}
This family of states is interesting to test the first bound since their entanglement of formation was numerically calculated by Audenaert {\it et al} \cite{audenaert} and was found to be very low. Figure (7) shows the bound provided by $R_{r}$ for the states $\rho' = e\rho(a) + (1 - e)(I/D)$. 

\begin{figure}
\begin{center}
\includegraphics[scale=0.45]{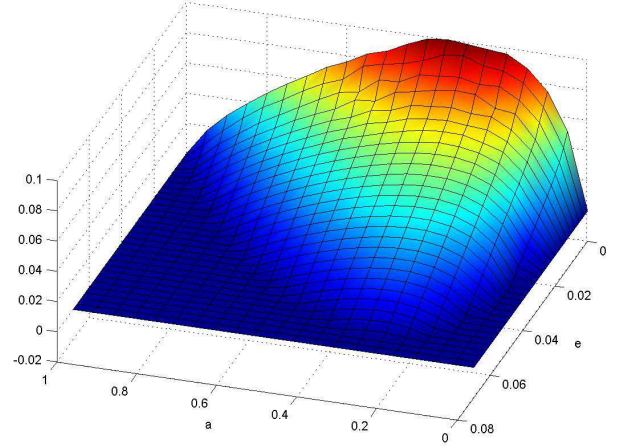}
\caption{(Coloronline) Lower bound for the Horodecki states using $R_{r}$.}
\end{center}
\end{figure}

For our second example we consider one of the unique states for which a analytic formula for $E_{F}$ is know. It was shown by Terhal and Vollbrecht \cite{terhal5} that for the isotropic states \cite{foot13}
\begin{equation*}
\rho_{F} = \frac{1 - F}{d^{2} - 1}\left( I - P^{+} \right) + FP^{+}
\end{equation*} 
\begin{equation}
E_{F}(\rho_{F}) = \frac{d\log(d - 1)}{d - 2}(F - 1) + \log(d), \hspace{0.2 cm} F \in \left[\frac{4(d - 1)}{d^{2}}, 1\right]
\end{equation}
From section (VII) we find
\begin{equation}
E_{F}(\rho_{F}) \geq (\log(d) - 1)\left(F  - \frac{1}{d} \right)
\end{equation}
We see that, in this case, for sufficiently large d, the difference of the bound and the actual value of $E_{F}$ is always less than F. 

Note that every entanglement witness $W$, after being normalized such that either $Tr(W) = 1$ or $W \leq I$ holds, can be used to deliver lower bounds to the entanglement of formation.

\section{Entanglement, Thermodynamics and Lattice Systems}

The study of entanglement properties of many-body systems, manly condensed matter, has received much attention recently \cite{nielsen2,arnesen,wang1,wang5,bose1,connor1,osborne1,osterloh,toth1,vidal6,ghose,brukner1}. Several important models have been analyzed and connections with thermodynamic variables, such as internal energy and magnetization, have been raised \cite{wang5,bose1,connor1,osborne1,toth1}. The negativity and concurrence have been the most used measures, partly due their easy computation, but also because they made possible the derivation of some interesting simple {\it thermodynamics like} equations. This can be understood from the view of the {\it witnessed entanglement}. Every quantie derived from (7) not only defines a measure of the degree of entanglement, but also gives a Hermitian operator, which vary for each state, whose expectation value quantifies the entanglement of the state in question. It is exactly the possibility of measure experimentally the amount of entanglement, which is a feature shared by all common thermodynamics variables, that makes quantities expressed by (7) useful to the study of entanglement thermodynamical properties.       
In this section we present, as an example, the study of entanglement in the XXX Heisenberg model with and without a magnetic field using $E_{\infty:1} = R_{G}$. The corresponding Hamiltonian is given by
\begin{equation}
H_{XXX} = J\sum_{i=1}^{N}\vec{\sigma}_{i}.\vec{\sigma}_{i+1} + B\sum_{i=1}^{N}\sigma_{i}^{z}.
\end{equation}                 
We first consider $B = 0$, in which case both Hamiltonians have $SU(2)$ symmetry. According to section (VII), we can restrict the EWs in (7) to the ones that also have $SU(2)$ symmetry. Then, from a standard result from representation theory \cite{eggeling1,weyl}, we find that all EWs with this symmetry can be written as
\begin{equation}
W = \sum_{i}\mu_{i}V_{i},
\end{equation}
where $V_{i}$ are unitary permutation operators. From analytic and numerical studies for the $XXX$ Heisenberg model of odd $N$ in the fundamental state and in the thermodynamical limit \cite{sakai}, we find that all other correlators are very small compared to the first neighbor correlators. We, thus, use the following ansatz for the optimal entanglement witness for the thermal states, at very low temperatures, of Hamiltonians (79) and (80) 

\begin{equation}  
W = \left(NI + \sum_{i=1}^{N}\left( \sigma_{i}^{x}\sigma_{i+1}^{x} + \sigma_{i}^{y}\sigma_{i+1}^{y} + \sigma_{i}^{z}\sigma_{i+1}^{z}\right)\right)/2N, 
\end{equation}     
where the factor 2$N$ in the denominator comes from $W \leq I$. Note that this is the EW introduced by Toth {\it et al} \cite{toth5}. Assuming that $|B| << |J|$, and using the continuity of OEWs, we find that for the XXX Heisenberg model, at temperatures sufficiently close to zero,
\begin{equation}
R_{G} \approx \frac{U - BM}{2NJ} - \frac{1}{2}, 
\end{equation}
where the magnetization and the internal energy are given, respectively, by $M = \sum_{i}\left<\sigma_{i}^{z}\right>$ and $U = \left<H\right>$. 

We now proceed analysing the relation between entanglement and the magnetic susceptibility ($\chi$) in thermal states of $H_{2}$. According to Brukner, Vedral and Zeilinger \cite{brukner1}, under temperatures close to zero and at zero external magnetic field, $\chi = (g^{2}\mu_{B}^{2}/kT)[NI + (1/3)\sum_{i}\vec{\sigma_{i}}.\vec{\sigma_{i+1}}]$. Thus,
\begin{equation}
\chi \approx \frac{2Ng^{2}\mu_{B}^{2}}{3kT} + \frac{2NR_{G}}{3kT}.
\end{equation}
Remarkably, we see that the susceptibility is given by a term which resembles the classical Curie law more a term which takes into account the entanglement presented in the state. The equation above can be seen as a quantitative version of the experimental result of Ghose {\it et al} \cite{ghose}, who shown that at very low temperatures the magnetic susceptibility of certain materials is affected by the existence of entanglement.  

\section{Conclusion}

Summarizing, we have presented a new perspective to the quantification of entanglement based on witness operators. Several important EMs were shown to fit into this scenario and a new infinite family of EMs was introduced. The usefulness of the {\it witnessed entanglement} was illustrated by the study of diverse features of entanglemnt, including super-selection rules constraints and efficiency of quantum information protocols. Finally, we have shown some interesting preliminary results in the study of thermodynamical properties of entanglement in macroscopic systems. 

We believe the results presented in this paper are only preliminary. The quantification on entanglement with EWs might be a very fruitful approach to development of the theory of entanglement, specially in the new applications of entanglement, such as in identifying quantum phase transitions and improving the approximation of mean field theories \cite{vedral5}.  
\section{acknowledgment}

The author would like to thank J. Eisert and D. Santos for helpful comments and especially Michal Horodecki for pointing out a mistake in a earlier version of this draft. Financial support from CNPQ is also acknowledged.


\begin{thebibliography}{99}
\bibitem{foot1} The compactness condition is required for preventing unbounded EW, which would made the minimization of equation (4) ill-defined.
\bibitem{foot2} Although \cite{lewenstein2} does not consider multipartite entanglement witnesses, the results presented there can be easily extended to the multipartite case.
\bibitem{foot3} We have performed the following change of variable from the matrix F considered in \cite{rains1}: W = I/d - F.
\bibitem{foot5} I should be interpreted as the limit of proper density matrices belonging to the Hilber space.
\bibitem{foot6} $P^{+} = \ket{\Psi^{+}}\bra{\Psi^{+}}$ is the projector on the improper state vector $\ket{\Psi^{+}} = \sum_{k \geq 0}\ket{k}_{A}\ket{k}_{B}$.
\bibitem{foot10} It should be clear that this result only holds if the group is composed of local unitaries.
\bibitem{foot7} Following \cite{vollbrecht1}, we are considering in this section only the bipartite case. Multipartite generalizations of the results presented are straightforward.
\bibitem{foot8} Equation (60) is valid only to the quanties that are entanglement monotones, such as $E_{n:m}$, ${\cal N}$, $C$ and $F_{d}$.
\bibitem{foot11} For m-entanglement, a witness operator is a Hermitian operator which is positive in all states of slater rank m.
\bibitem{foot9} A twirling operation is obtained by Alice applying an arbitrary unitary U, by Bob applying $U^{*}$, and by then deleting the classical information concerning which unitary has been applied \cite{vidal3}.
\bibitem{foot13} In this section we are using the same parametrization of the isotropic states used in \cite{terhal5}, different from the one of section VIII.
\end{thebibliography}
\end{document}